\documentclass{article}
\usepackage{placeins}

\usepackage{xspace}

\def\showdraftbox{0}
\def\showauthornotes{0}

\usepackage{color}
\definecolor{ForestGreen}{rgb}{0.1333,0.5451,0.1333}
\usepackage[letterpaper,
            colorlinks,linkcolor=ForestGreen,citecolor=ForestGreen,
            backref=page,
            bookmarks,bookmarksopen,bookmarksnumbered]
           {hyperref}

\usepackage{cleveref}

\crefname{eq}{equation}{equations}
\crefname{thm}{Theorem}{Theorems}

\usepackage{amstext,amsfonts,bbm,algorithm,algorithmicx,xspace,nicefrac}
\usepackage{amsmath,amssymb}    % AMS
\usepackage{amsthm}
\usepackage{thmtools}
\usepackage{thm-restate}
\usepackage{color,stmaryrd,enumerate,latexsym,bm,amsfonts,
  wrapfig,verbatim,textcomp}
\usepackage[small]{caption}
\usepackage{comment} 
\usepackage{epsfig} 
\usepackage{tabularx}
\usepackage{tabu}
\usepackage{fancybox,graphicx,url}
\usepackage[inline]{enumitem}
\usepackage{fullpage}
\usepackage{booktabs}
\usepackage{commath}
\usepackage{tablefootnote}

\newcommand{\defeq}{\stackrel{\textup{def}}{=}}

\newtheorem{theorem}{Theorem}[section]
\newtheorem{corollary}[theorem]{Corollary}
\newtheorem{lemma}[theorem]{Lemma}

\newtheorem{fact}[theorem]{Fact}

\theoremstyle{definition}
\newtheorem{Definition}[theorem]{Definition}

\def\abs#1{\left| #1 \right|}
\renewcommand{\norm}[1]{\ensuremath{\left\lVert #1 \right\rVert}}

\newcommand\rea{\mathbb R}

\newcommand{\marginlabel}[1]%
{\mbox{}\marginpar{\it{\raggedleft\hspace{0pt}#1}}}
\newcommand{\poly}{\mathrm{poly}}
\DeclareMathOperator{\polylog}{polylog}
\definecolor{Mygray}{gray}{0.8}

 \ifcsname ifcommentflag\endcsname\else
  \expandafter\let\csname ifcommentflag\expandafter\endcsname
                  \csname iffalse\endcsname
\fi

\ifnum\showauthornotes=1

\else

\fi

\ifnum\showauthornotes=1
\newcommand{\Authornote}[2]{{\sf\color{red}{[#1: #2]}}}
\newcommand{\Authoredit}[2]{{\sf\color{red}{[#1]}\color{blue}{#2}}}
\newcommand{\Authorcomment}[2]{{\sf \color{gray}{[#1: #2]}}}
\newcommand{\Authorfnote}[2]{\footnote{\color{red}{#1: #2}}}
\newcommand{\Authorfixme}[1]{\Authornote{#1}{\textbf{??}}}
\newcommand{\Authormarginmark}[1]{\marginpar{\textcolor{red}{\fbox{%\Large
#1:!}}}}
\else
\newcommand{\Authornote}[2]{}
\newcommand{\Authoredit}[2]{}
\newcommand{\Authorcomment}[2]{}
\newcommand{\Authorfnote}[2]{}
\newcommand{\Authorfixme}[1]{}
\newcommand{\Authormarginmark}[1]{}
\fi

\newenvironment{fminipage}%
  {\begin{Sbox}\begin{minipage}}%
  {\end{minipage}\end{Sbox}\fbox{\TheSbox}}

\newlength{\pgmtab}  %  \pgmtab is the width of each tab in the
 {
	\begin{enumerate}}{\end{enumerate}}

\def\qedsketch{\ifmmode\Box\else{\unskip\nobreak\hfil
\penalty50\hskip1em\null\nobreak\hfil$\Box$
\parfillskip=0pt\finalhyphendemerits=0\endgraf}\fi}

\newlength{\tpush}
\newcommand{\handout}[5]{
   \noindent
   \begin{center}
   \framebox{ \vbox{ \hbox to \textwidth { {\bf \coursenum\ :\  \coursename} \hfill #5 }
       \vspace{3mm}
       \hbox to \textwidth { {\Large \hfill #2  \hfill} }
       \vspace{1mm}
       \hbox to \textwidth { {\it #3 \hfill #4} }
     }
   }
   \end{center}
   \vspace*{4mm}
   \newcommand{\lecturenum}{#1}
   \addcontentsline{toc}{chapter}{Lecture #1 -- #2}
}

\ifnum\showdraftbox=1

\else

\fi

\newcommand{\Oh}{\ensuremath{\mathcal{O}}}

\newcommand{\bv}[1]{\mathbf{#1}}
\newcommand{\expct}[1]{\ensuremath{\mathbb{E}\left[#1\right]}}
\newcommand{\prob}[1]{\ensuremath{\mathop{\text{\normalfont \textbf{P}}}}\left[#1\right]}
\newcommand{\expctt}[2]{\ensuremath{\mathbb{E}_{#1}\left[#2\right]}}
\newcommand{\Proof}[0]{\smallskip\noindent\textit{\textbf{Proof:}}\quad}

\newcommand{\Proofof}[1]{\smallskip\noindent\textit{\textbf{Proof of #1:}}\quad}

\newcommand{\QED}[0]{\hfill\ensuremath{\blacksquare}\medspace\\}
\DeclareMathOperator{\Exp}{Exp}
\newcommand\given[1][]{\:#1\vert\:}

\newcommand{\streamSparsify}{\textsc{StreamSparsify}\xspace}

\allowdisplaybreaks[4]

\newcommand{\eps}{\epsilon}

\usepackage{caption,mdframed,setspace}

\newlength{\algtopspace}
\newlength{\algpostcaptionspace}
\makeatletter

\newenvironment{shiftedflalign*}{%
    \start@align\tw@\st@rredtrue\m@ne
    \hspace{1.5 em}
}{%
    \endalign
}

\title{
A Framework for Analyzing Resparsification Algorithms
}

\author{
  Rasmus Kyng\\
Yale University
\thanks{rasmus.kyng@yale.edu}
\and
Jakub Pachocki\\
Harvard University
\thanks{meret@seas.harvard.edu}
\and
Richard Peng\\
Georgia Tech
\thanks{rpeng@cc.gatech.edu}
\and 
Sushant Sachdeva\\
Google
\thanks{sachdevasushant@gmail.com}
}

\date{}

\begin{document}

\maketitle

\begin{abstract}
  A spectral sparsifier of a graph $G$ is a sparser
  graph $H$ that approximately preserves the quadratic form of $G$, i.e.
  for all vectors $x$, $x^T L_G x \approx x^T L_H x$, where $L_G$ and $L_H$
  denote the respective graph Laplacians. Spectral sparsifiers
  generalize cut sparsifiers, and have found many applications in
  designing graph algorithms. In recent years, there has been interest
  in computing spectral sparsifiers in semi-streaming and dynamic
  settings.
  Natural algorithms in these settings often involve repeated
  sparsification of a graph, and in turn accumulation of errors
  across these steps.
  We present a framework for analyzing algorithms that perform repeated
  sparsifications that only incur error corresponding to a single
  sparsification step, leading to better results for many of these
  resparsification-based algorithms.

  As an application, we show how to maintain a spectral sparsifier
  in the semi-streaming setting: We present a simple algorithm that,
  for a graph $G$ on $n$ vertices and $m$ edges, 
  computes a spectral sparsifier of $G$ with $\Oh(n \log n)$
  edges in a single pass over $G$, using only $\Oh(n \log n)$ space, and
  $\Oh(m \log^2 n)$ total time.
  This improves on previous best semi-streaming algorithms for
  both spectral and cut sparsifiers by a factor of $\log{n}$
  in both space and runtime.
  The algorithm extends to semi-streaming row sampling for
  general PSD matrices. 
 We also use our framework to combine a spectral
 sparsification algorithm by Koutis with improved spanner
 constructions to give a parallel algorithm for constructing
 $\Oh(n\log^2{n}\log\log{n})$ sized spectral
 sparsifiers in $\Oh(m\log^2{n}\log\log{n})$ time.
 This is the best combinatorial graph sparsification algorithm
 to date, and the size of the sparsifiers produced is only
 a factor $\log{n}\log\log{n}$ more than ones produced by numerical routines.

\end{abstract}

\section{Introduction}

\paragraph{Graph Sparsifiers.}
Consider an undirected graph $G$ with vertices $V,$ and edges $E,$
such that $|V|=n$ and $|E|=m.$
A sparsifier of $G$ is a, hopefully sparser, graph $H$ that
approximates $G$ in a meaningful way. Bencz\'{u}r and
Karger~\cite{BenczurK96} introduced the notion of a combinatorial
sparsifier (also called a cut-sparsifier), where $H$ is also a graph
on $V$ such that for every cut of $V,$ its value in $H$ is within a
$1 \pm \eps$ factor of the same cut in $G$.
Their result~\cite{BenczurK96} gave an algorithm to construct a
cut-sparsifier $H$ with $\Oh(n\eps^{-2} \log n)$ edges in
$\Oh(m \log^3 n)$ time, and a better running time of
$\Oh(n\log^2{n} + m)$ was shown in~\cite{FungHHP11}.
Cut sparsifiers led to faster algorithms for approximating $s$-$t$
min-cuts~\cite{BenczurK96}, sparsest cuts~\cite{BenczurK96,
  KhandekarRV06}, and undirected maximum-flows~\cite{KargerL15}.

Spielman and Teng~\cite{SpielmanT04} generalized the notion of 
graph sparsification to the spectral setting.
We define the Laplacian of $G$ to be the unique symmetric matrix
$L_{G}$ such that for all vectors $x \in \rea^{n},$ we have
$x^{\top} L_{G} x = \sum_{(i,j) \in E} (x_u - x_v)^2,$ the natural
quadratic form on $G.$
$H$ is an $\eps$-spectral-sparsifier of $G$ if it approximately
preserves the natural quadratic form, \emph{i.e.}, for all
$x \in \rea^{V},$ we have
$(1-\eps) x^{\top} L_{G} x \le x^{\top} L_{H} x \le (1+\eps) x^{\top}
L_{G} x.$
By considering $x \in \{0,1\}^{n},$ it is immediate that a
spectral-sparsifier is also a cut-sparsifier.  Spielman and
Teng~\cite{SpielmanT04} gave an algorithm for constructing spectral
sparsifiers with $\Oh(n\eps^{-2} \log^{\Oh(1)} n)$ edges in
$\Oh(m \log^{\Oh(1)} n)$ time, and utilized them to design
a nearly-linear time solver for linear systems in graph
Laplacians.
These nearly-linear time solvers have been used as
primitives in the design of fast algorithms~\cite{KelnerM09,
  OrecchiaSV12, KelnerMillerPeng, KoutisLP15, KyngRS15}.
Spielman and Srivastava~\cite{SpielmanS08} showed that sampling edges
proportional to their leverage scores and rescaling them
appropriately, produces a spectral sparsifier with $\Oh(n\eps^{-2} \log
n)$ edges with high probability.

The above works~\cite{KoutisLP15,LeeS15,AndoniCKQWZ16}
were motivated by the desire to develop better sparsification
algorithms.
Progress on
sampling-based constructions include more robust sampling
methods steps~\cite{KelnerL13}, and faster estimations of sampling
probabilities~\cite{KoutisLP15}.
When combined with improved solvers for linear systems
in graph Laplacians~\cite{KoutisMP11,Peng13:thesis,KyngLPSS16},
these works lead to the current state-of-the-art running time of
$\Oh(m \log^2 n)$ for finding spectral sparsifiers with
$\Oh(n  \epsilon^{-2} \log{n})$ edges.

On the other hand, when such routines are used to analyze
large graphs, there appears to be a significant gap between 
linear system solving oriented approaches motivated by
these theoretical studies~\cite{MavroforakisGKT15},
and the local, often combinatorial approaches taken
in many practical studies~\cite{JohnS16:arxiv,KrishnanFS13}.
In this paper, we build upon recent developments in local
steps for solving linear systems~\cite{KyngS16} to address
a major obstacle for analyzing these local sparsification
routines: the accumulation of error across randomized steps.
We give a systematic way to analyze the error accumulation, and
a versatile sufficient condition for showing that these
local steps are no worse than global sampling steps.
This result has direct consequences for sparsification in
semi-streaming settings, as well as combinatorial constructions
of graph sparsifiers.

\paragraph{Semi-streaming algorithms.}
For analyzing massive graphs, it is often prohibitive to even store
the entire graph in memory. One model for describing such graphs is 
the streaming computational model, where the graph is presented as
a stream of edges, and the algorithm is limited to a few passes over
the stream, and space that is polylogarithmic in the input size. This
model turns out to be too restrictive for many graph problems, as even
simple problems such as $s$-$t$ connectivity require $\Omega(n/k)$
space with $k$ passes~\cite{HenzingerRS99}. The semi-streaming model, where the algorithm is
permitted $\Oh(n \polylog n)$ space has been more fruitful for designing
graph algorithms~\cite{FeigenbaumKMMSZ05}.

The problem of constructing graph sparsifiers in the semi-streaming
model was first studied by Ahn and Guha~\cite{AhnG09}.
They construct a cut-sparsifier for $G$ with
$\Oh(n\eps^{-2} \log n \log \frac{m}{n})$
edges using $\Oh(n\eps^{-2} \log n \log \frac{m}{n})$ space.
Kelner and Levin~\cite{KelnerL13} gave a simple single-pass
algorithm for constructing a spectral sparsifier.
This algorithm maintains a sparsifier at every step by adding
incoming edges to it, and resparsifying once the edge count
reaches a certain threshold.
The immediate concern is that we can not compute the
sampling probabilities with respect to the final graph.
The key idea in~\cite{KelnerL13} is to compute the sampling
probabilities with respect to the current sparsifier
and perform rejection sampling using the ratio between these and the
earlier probabilities.
Assuming we have a good sparsifier of the graph that we have seen so far,
we can obtain upper bounds for sampling probabilities for the final graph.
Unfortunately, this introduces dependencies between the sampled edges,
and the argument in~\cite{KelnerL13} has difficulties handling
these dependencies (this is discussed in more detail in~\cite{CohenMP16}).
Our algorithm is closely related to that of Kelner and Levin, but only
works with sampling probabilities computed based on the current graph.

\subsection{Our Results : Analyzing Resparsification.}
There are two main challenges in analyzing resparsification routines such as the one from~\cite{KelnerL13}.
The first is the dependencies between the edge samples.
These dependencies prevent us from analyzing the entire sampling
process as a single sparsification step, or invoke matrix
concentration inequalities.
The second is understanding the accumulation of errors:
if we resparsify a graph $k$ times with an error of $1\pm \eps$ at each step,
we are only guaranteed an $(1\pm \eps)^{k} \approx 1 \pm k\eps$ approximation
to the original graph by the end.
However, if the resparsifications are independent, an
adversarial accumulation of error seems too pessimistic. 

In this paper, we present a framework for analyzing resparsification
routines, that allows us to handle the dependencies and avoids
pessimistic accumulation of error, resulting in improved
parameters. Applying our framework, we obtain the following result in
the semi-streaming setting, for an algorithm very similar to the one
in~\cite{KelnerL13}.
\footnote{The primary difference between the algorithms is that we decide whether to keep each edge independently, while the algorithm in~\cite{KelnerL13} partially employs sampling with replacement.}
\begin{theorem}[Semi-Streaming Sparsification]
\label[thm]{thm:streaming}
  For all $\eps > 0,$ the algorithm \streamSparsify, given as input
  a graph $G$ with $n$ vertices and $m$ edges, outputs a graph $H$
  such that $H$ is an $(1\pm\eps)$-spectral sparsifier of $G$ with probability
  $1-\frac{1}{\poly(n)}.$ {\streamSparsify} requires just one
  pass over $G,$ and runs in $\Oh(m \log^2 n)$ total time using 
  $\Oh(n\eps^{-2} \log n)$ space.
\end{theorem} 
This result can viewed as reducing the
  memory usage of sparsification by leverage scores~\cite{SpielmanS08}
  to the size of its output, $\Oh(n\eps^{-2} \log n)$, without any
  additional caveats. In the semi-streaming setting, it gives a tight
  analysis of the spectral sparsifier construction proposed
  in~\cite{KelnerL13}, and also improves on the previous best constructions of
  cut sparsifiers~\cite{AhnG09}. The fact that the error due to
  resparsifications does not accumulate is a priori
  surprising!

As another application of our framework, we show that it results in a
better analysis of an algorithm of Koutis for constructing sparsifiers
in parallel~\cite{Koutis14}. Combining with a better sparsifier
construction, we obtain the following result:
\begin{theorem}
\label[thm]{thm:parallel-sparsifier}
For all $\epsilon > 0$, the algorithm \textsc{ParallelSparsify} given
as input a graph $G$ with $n$ vertices and $m$ edges, outputs a graph
$H$ with $\Oh(n \eps^{-2} \log^2 n \log \log n)$ edges such that $H$
is a $(1\pm\epsilon)$-spectral sparsifier of $G$ with probability at
least $1-\frac{1}{\poly(n)}$.  The total work of the algorithm
$\Oh(m \eps^{-2} \log^2 n \log \log n)$ and the depth is
$\Oh(\eps^{-2} \log^{4} n \log^{*} n)$.
\end{theorem}
This result gives the best combinatorial construction of spectral sparsifiers,
improving over the work of Kapralov and Panigrahy~\cite{KapralovP12} that
constructs a sparsifier with $\Oh(n\eps^{-3} \log^3 n)$ edges in
$\Oh(m\eps^{-3} \log^3 n)$ time.
It also improves on the previous best result for parallel
  construction of spectral sparsifiers due to Koutis~\cite{Koutis14}
  that constructs a sparsifier with $\Oh(n \eps^{-2} \log^{6} n)$
  edges using $\Oh(m \eps^{-2} \log^5 n)$ work and
  $\Oh(\eps^{-2} \log^{6} n)$ depth.
  A detailed comparison to the other sparsification algorithms is
  in Figure~\ref{fig:previous} of Section~\ref{sec:sparsify}.

The question of whether
similar or better guarantees can be obtained using combinatorial
graph algorithms is important for deciding how graph
sparsifiers can be used as an algorithmic tool.
Works in this direction have centered around using sparsifiers
either as direct effective resistance estimators~\cite{KapralovP12}
or as part of resparsification schemes~\cite{Koutis14}.

\paragraph{Framework and Techniques.} 
Our framework for analyzing resparsification algorithms has several
key components.  We describe these components here with the
semi-streaming algorithm as our running example.
Our algorithm can be described roughly as follows: Let
$S = \Oh(n \eps^{-2} \log n)$ be the target size of our sparsifier. We
start with the empty graph, and keep adding incoming edges until the
sparsifier has size $S.$ We then use the current sparsifier to
estimate the leverage scores of the edges, and use them to sparsify
the current sparsifier down to size $S/2,$ and continue.

The first key conceptual idea is that we formulate the entire
algorithm as a matrix martingale. In the semi-streaming setting, this
corresponds to studying the Laplacian of the current sparsifier
\emph{plus the remaining edges in the stream}. Three simple
observations illustrate the usefulness of this view:
1. Initially, this matrix is the Laplacian of the whole graph,
2. At the end, this matrix is the Laplacian of the sparsifier
outputted by the algorithm, and
3. Each sparsification step preserves the expectation of this matrix.

The key tool we use for analyzing the matrix martingale is Freedman's
inequality~\cite{Tropp11}. Freedman's inequality is the martingale
counterpart to Bernstein's inequality.
It allows us to control the deviation in the martingale via bounding
the predictable quadratic variation, which is roughly the sum of the
variances over the steps.
In our applications, this predictable quadratic variation is then
easily bounded using matrix Chernoff bounds~\cite{Tropp12}.

Our improved constructions of spectral sparsifiers in Section~\ref{sec:sparsify}
are based on analyzing the resparsification steps of a
sparsification algorithm due to Koutis~\cite{Koutis14}
in our framework.
This routine uses bundles of spanners to repeatedly
compute upper bounds of effective resistances that suffice
for reducing edge count by a constant factor.
We also provide improved parallel algorithms for finding these
estimates in Section~\ref{subsec:spannerImproved}.

%%% These are comments needed for working with emacs
%%% Local Variables: 
%%% mode: latex
%%% TeX-master: "resparsify"
%%% End: 

\section{Preliminaries}
For symmetric matrices, we write $\bv{A} \preceq \bv{B}$ iff
$\bv{B}-\bv{A}$ is a positive-semidefinite matrix.
Throughout the paper, we use `with high probability' for events that
happen with probability at least $1 - \frac{1}{\poly(n)}$, where
$\poly(n)$ can be set to an arbitrarily large polynomial by only adjusting constant factors.
\paragraph{Laplacians and Sparsifiers.}
We consider connected undirected graphs $G =(V,E)$, with vertices $V$
and edges $E.$ We assume the edges have positive weights
$w : E \to \rea_{+}.$  Let $n = \abs{V}$ and $m = \abs{E}$.
Let $\bv{e}_{i}$ denote the $i^{\textrm{th}}$ standard basis vector
with a 1 in the $i^{\textrm{th}}$-coordinate and 0 otherwise. For
every edge $e,$ we assign an arbitrary order to its endpoints, and for
$e = (u,v),$ we define $\bv{b}_{e} = \bv{e}_{u} - \bv{e}_{v}.$
Finally, we define the Laplacian of $G$ as
$\bv{L}_{G} = \sum_{e \in E} w(e) \bv{b}_{e} \bv{b}_{e}^{\top}.$ For every
$x \in \rea^{n},$ we have,
\begin{align*}
x^{\top} \bv{L}_{G} x
 = \sum_{e \in E} w(e) x^{\top}\bv{b}_{e}
\bv{b}_{e}^{\top}x
= \sum_{e=(u,v)\in E} w(e) (x_{u}-x_{v})^2.
\end{align*}
It is immediate that $\bv{L}_{G} \succeq \bv{0}.$ Note that
$\bv{L}_{G}$ is independent of the choice of the ordering for each
edge.

\begin{Definition}[Sparsifier]
  A graph $H(V,E^{\prime})$ is said to be a
  $(1\pm\eps)$-spectral-sparsifier of $G(V,E)$ if we have
  $E^{\prime} \subseteq E,$ and for all $x \in \rea^{V},$
\[ 
  (1-\eps) x^{\top} \bv{L}_{G} x \le x^{\top} \bv{L}_{H} x \le
  (1+\eps) x^{\top} \bv{L}_{G} x.\]
\end{Definition}

\paragraph{Matrix Concentration.}
We will use the following two theorems due to Tropp
\cite{Tropp11,Tropp12}: 
\begin{theorem}[Matrix Chernoff]
    \label[thm]{thm:chernoff}
    Consider a finite sequence $\{\bv{X}_k\}$ of independent, random,
    symmetric  $ n \times n$-matrices.
    Assume that each random matrix satisfies
    \begin{align*}
        \bv{X}_k \succeq \bv{0} \mbox{ and } \norm{\bv{X}_k} \leq R \mbox{ almost surely.}
    \end{align*}
    Define
    \begin{align*}
        \mu_{\max} \defeq \norm{\sum_k \expct{\bv{X}_k}}.
    \end{align*}
    Then, we have, for every $\delta > 0,$
    \begin{align*}
        \prob{\norm{\sum_k \bv{X}_k} \geq (1+\delta)\mu_{\max}}
		\leq n\cdot \left(\frac{e^\delta}{(1+\delta)^{1+\delta}}\right)^{\mu_{\max}/R}.
    \end{align*}
\end{theorem}

Recall, a set of random variables
$\bv{Y}_0, \bv{Y}_1, \bv{Y}_2, \ldots$ that take
values over symmetric $n \times n$ matrices is said to be a matrix
martingale, if (informally), each $\bv{Y}_{j}$ only depends on the
previous variables $\bv{Y}_{0},\ldots,\bv{Y}_{j-1}$ and
$\expctt{j-1}{\bv{Y}_{j}} = \bv{Y}_{j-1},$ where $\expctt{j-1}{\cdot}$
denotes expectation conditional on $\bv{Y}_{0},\ldots,\bv{Y}_{j-1}$.
\begin{theorem}[Matrix Freedman]
    \label[thm]{thm:freedman}
    Let $\bv{Y}_0, \bv{Y}_1, \bv{Y}_2, \ldots$ be a matrix martingale
    whose values are symmetric $n \times n$ matrices, and let
    $\bv{X}_1, \bv{X}_2, \ldots$ be the difference sequence
    $\bv{X}_{i} = \bv{Y}_{i} - \bv{Y}_{i-1}$.  Assume that the
    difference sequence is uniformly bounded in the sense that
    \begin{align*}
        \norm{\bv{X}_k} &\leq R\mbox{ almost surely, for all } k.
    \end{align*}
    Define the predictable quadratic variation process of the martingale:
    \begin{align*}
        \bv{W}_k \defeq \sum_{j=1}^k \expctt{j - 1}{\bv{X}_j^2}\mbox{, for all } k.
    \end{align*}
    Then, for all $t > 0$ and $\sigma^2 > 0$,
    \begin{align*}
        &\prob{\exists k : \norm{\bv{Y}_k} \geq t \mbox{ and } \norm{\bv{W}_k} \leq \sigma^2}
		\leq n\cdot\exp\left(-\frac{-t^2/2}{\sigma^2+Rt/3}\right).
    \end{align*}
\end{theorem}

%%% These are comments needed for working with emacs
%%% Local Variables:
%%% mode: latex
%%% TeX-master: "resparsify"
%%% End:

\section{Resparsification Game}
\label{sec:game}
We introduce a resparsification game that is an abstraction of a
large class of resparsification algorithms. We will make minimal
assumptions about the algorithm (or adversary) and prove that with the
right choice of parameters, resparsification does not lead to error
accumulation.
In the next few sections, we will present algorithms for
semi-streaming sparsification and parallel sparsification, and prove
that they satisfy the assumptions of our resparsification game.

\paragraph{The Game.} Say we're given $\epsilon \in \left(0, \frac{1}{2}\right),$ and $m$
vectors $\bv{a}_1,\ldots,\bv{a}_m \in \mathbb{R}^n$, such that
\begin{align*}
    \sum_{i=1}^m \bv{a}_i \bv{a}_i^\top = \bv{M}.
\end{align*}
Set $\alpha := \Oh (  \log n \epsilon^{-2})$,
and initialize $w_i$ to $1$ for each $1 \leq i \leq m$.
We will analyze a game played by an adversary on the
weights $w_i$.
The game consists of a single move, repeated while
the game is not over:
\begin{enumerate}
    \item The adversary picks any $i \in \{1, \ldots, m\}$ and $p \in
    (0, 1]$ such that $w_i \neq 0$ and $\frac{w_i}{p} \bv{a}_i^\top \bv{M}^\dagger
\bv{a}_i \leq \frac{1}{\alpha}$\footnotemark. If there's no such pair, the game
      ends and the adversary loses.
    \item With probability $p$, set $w_i \leftarrow \frac{w_i}{p}$; otherwise, set $w_i \leftarrow 0$.
\end{enumerate}
\footnotetext{Here, $\bv{M}^\dagger$ denotes the Moore-Penrose pseudoinverse of $\bv{M}$.}

The adversary wins if at some point in the game, the matrix
$\sum w_i \bv{a}_i \bv{a}_i^\top$ is not a
$(1 \pm \epsilon)$-approximation to $\bv{M}$. More formally, the
adversary wins if at any point in the game, the following condition fails to hold:
\[ (1-\eps) \bv{M} \preceq \sum w_i \bv{a}_i \bv{a}_i^\top \preceq
(1+\eps) \bv{M}.\]

Note that the main power gained by the adversary in this
setting compared to static graph sparsification
is that it can pick $i$ and $p$ based on the matrices
and weights chosen so far.
We will show that with appropriate settings of constants
in $\alpha$, the probability that the adversary wins
the game can be controlled at $\Oh(n^{-c})$.

\begin{theorem}
\label[thm]{thm:main}
    With high probability, the adversary will not win
	the game defined above.
\end{theorem}

For the analysis, we will make the following assumptions without loss of generality:

\begin{itemize}
    \item $\bv{M} = \bv{I}$. The case of arbitrary $\bv{M}$ can be easily reduced to this case by multiplying all $\bv{a}_i$ by $\bv{M}^{\dagger/2}$ and projecting out the space corresponding to $\ker(\bv{M})$. This is a standard reduction, see e.g. \cite{BSS}.
    \item For all $i$, $\bv{a}_i^\top \bv{a}_i \leq \frac{1}{\alpha}$.
        Any rows with higher leverage score can simply be excluded from the game.
    \item The adversary can only pick $p \geq \frac{1}{2}$.
        Note that the adversary can still simulate an arbitrarily small value of $p$ through multiple moves.
\end{itemize}

We will also assume without loss of generality that the randomness in the game is generated in the following way.
Let $x_1, \ldots, x_m \sim \Exp(1)$ be independent random variables drawn from the exponential distribution with parameter $1$.
Whenever the adversary picks a pair $(i, p)$, the weight $w_i$ is not set to zero iff
\begin{align*}
    \frac{w_i}{p} \leq e^{x_i}.
\end{align*}
The probability of this event, conditioning on the history of the game, is
\begin{align*}
    \prob{e^{x_i} \geq \frac{w_i}{p} \given[\Big] e^{x_i} \geq w_i} = p
\end{align*}
because of the memoryless property of the exponential distribution.

\subsection{Bounding the Predictable Quadratic Variation}
We consider a martingale with the difference $\bv{X}_j$
corresponding to the $j$-th move by the adversary, or
$\bv{0}$ if the adversary has made fewer than $j$ moves.
Assume the adversary chooses row $i$ and keeping
probability $p$; then we have
\begin{align*}
    \bv{X}_j \defeq 
    \begin{cases}
        \frac{1-p}{p} \cdot w_i \bv{a}_i\bv{a}_i^\top &\mbox{ with probability $p$} \vspace{0.2cm}\\
        -w_i \bv{a}_{i}\bv{a}_i^\top &\mbox{ otherwise.}\\
    \end{cases}
\end{align*}
Let $\{\bv{W}_k\}$ be the predictable quadratic variation process of the martingale given by the $\{\bv{X}_j\}$:
\begin{align*}
    \bv{W}_k \defeq \sum_{j=1}^k \expctt{j-1}{\bv{X}_j^2}.
\end{align*}
In order to bound $\norm{\bv{W}_k},$ we need an auxiliary lemma. For all $i \in {1, \ldots, m}$ define
\begin{align*}
    \overline{w}_i \defeq \min\left(e^{x_i}, \frac{1}{\alpha \bv{a}_i^\top\bv{a}_i}\right).
\end{align*}
Note that the $\overline{w}_i$ are independent random variables, and,
throughout the entire game, we have, $w_i \leq \overline{w}_i$.
\begin{lemma}
    With high probability, we have that
\Authornote{Richard}{I got rid of the norms since
this is invoked with $\preceq$}
    \begin{align*}
		\sum_{i=1}^{m} \overline{w}_i^2 (\bv{a}_i \bv{a}_i^\top)^2
		\preceq  \frac{4}{\alpha} \bv{I}.
    \end{align*}
    \label{lem:gamebound}
\end{lemma}
\Proof
Let $\bv{Z}_i \defeq \overline{w}_i^2 (\bv{a}_i\bv{a}_i^\top)^2$, for $i = 1, \ldots, m$.
Our goal is to bound the norm of the sum of the independent matrices $\bv{Z}_i$.
First of all, note that we always have $\norm{\bv{Z}_i} \leq \frac{1}{\alpha^2}$.
Moreover,
% \begin{align*}
\begin{equation*}
\begin{aligned}
    \expct{\bv{Z}_i} &= \left(\prob{e^{x_i} > \frac{1}{\alpha
          \bv{a}_i^\top\bv{a}_i}} \left(\frac{1}{\alpha
          \bv{a}_i^\top\bv{a}_i}\right)^2
      + \int_0^{-\log \alpha \bv{a}_i^\top\bv{a}_i} e^{-x} (e^x)^2 dx\right) \cdot (\bv{a}_i\bv{a}_i^\top)^2\\
    &= \left(\frac{2}{\alpha \bv{a}_i^\top\bv{a}_i} - 1\right) \cdot (\bv{a}_i\bv{a}_i^\top)^2
                       \preceq \frac{2}{\alpha} \cdot \bv{a}_i\bv{a}_i^\top.
% \end{align*}
\end{aligned}
\end{equation*}
Thus, we get $\sum_{i=1}^m \expct{\bv{Z}_i} \preceq
\frac{2}{\alpha} \bv{I}$.
\Cref{thm:chernoff} with $\mu_{\max} =
\frac{2}{\alpha}, \delta = \frac{2}{\alpha\cdot\mu_{\max}}, R = \frac{1}{\alpha^2}$ then gives the lemma.
\QED

We can now bound the total quadratic variation
$\bv{W}_{k} = \sum_{j=1}^k \expctt{j - 1}{\bv{X}_j^2}$
as specified in the Matrix Freedman inequality
in Theorem~\ref{thm:freedman}.
\begin{lemma}
\label{lem:loww}
    With high probability, for all $k$ we have that
    \begin{align*}
        \norm{\bv{W}_k} \leq \frac{16}{\alpha}.
    \end{align*}
\end{lemma}
\Proof
Suppose the adversary picks $i$ and $p$ in the $j$-th move.
In what follows, $w_i$ refers to the value of $w_i$ 
in the $j$-th move.
We have
\[
\expctt{j-1}{\bv{X}_j^2}
	= \frac{1-p}{p} \cdot w_i^2 (\bv{a}_{i}\bv{a}_i^\top)^2
	= \left(\frac{w_i^2}{p} - w_i^2\right) \left(\bv{a}_{i}\bv{a}_i^\top\right)^2
	 \preceq \left(\frac{w_i^2}{p^2} - w_i^2\right)
      \left(\bv{a}_{i}\bv{a}_i^\top\right)^2.
\]
In order to bound $\bv{W}_k,$ we sum the above expression for all $j,$
grouping them by the index $i$ picked in the $j^{\textrm{th}}$
round. For every fixed $i,$ we have a telescoping sum. 
Since $p \le \frac{1}{2},$ and $w_{i}$ is always less than
$\overline{w}_{i},$ the sum for $i$ is spectrally upper bounded by
$4 \overline{w}_i^2 (\bv{a}_i\bv{a}_i^\top)^2$.  Hence, we have
\[
    \bv{W}_k
	\preceq \sum_{i=1}^m 4 \cdot \overline{w}_i^2 (\bv{a}_i\bv{a}_i^\top)^2.
\]
The claim now follows from \Cref{lem:gamebound}.. 
\QED

\Proofof{\Cref{thm:main}}
The condition on $p_i$ ensure
that $\norm{\bv{X}_k} \leq \frac{1}{\alpha}$.
Together with \Cref{lem:loww}, applying \Cref{thm:freedman} with $t = \epsilon, \sigma^2 = \frac{16}{\alpha}, R = \frac{1}{\alpha}$ yields the theorem.
\QED

\subsection{Application to Streaming Sparsification}
\label{sec:application}

\begin{figure}[ht]
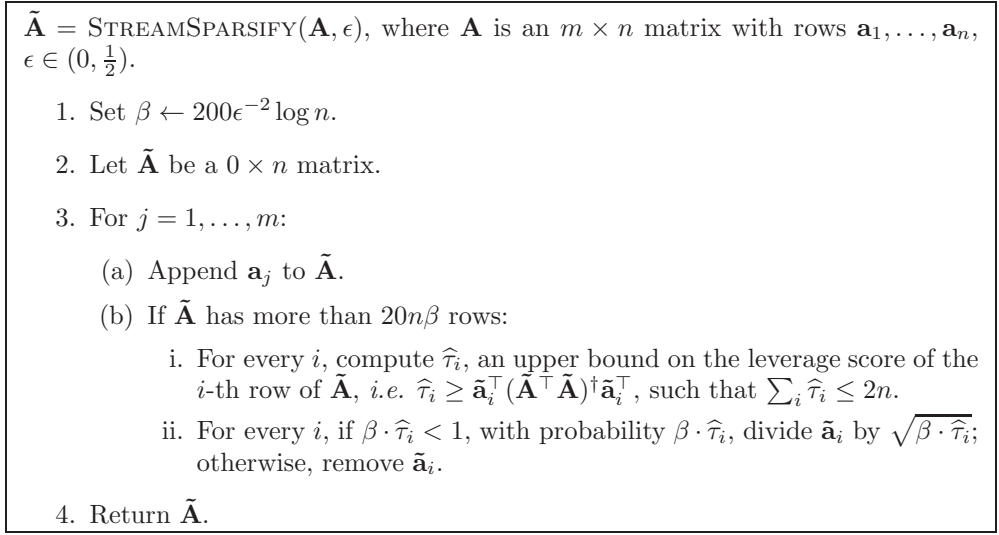

\noindent
\centering
\fbox{
\begin{minipage}{5in}
    \noindent $\bv{\tilde{A}} = \streamSparsify (\bv{A}, \epsilon)$,
    where $\bv{A}$ is an $m \times n$ matrix with rows $\bv{a}_1, \ldots, \bv{a}_n$, $\epsilon \in (0, \frac{1}{2})$.
\begin{enumerate}
\item Set $\beta \leftarrow 200 \epsilon^{-2} \log n$.
\item Let $\bv{\tilde{A}}$ be a $0 \times n$ matrix.
\item For $j = 1, \ldots, m$:
    \begin{enumerate}
        \item Append $\bv{a}_j$ to $\bv{\tilde{A}}$.

        \item If $\bv{\tilde{A}}$ has more than $20 n \beta$ rows:
            \begin{enumerate}
            \item\label{step:lev} For every $i,$ compute $\widehat{\tau}_{i},$ an
              upper bound on the leverage score of the $i$-th row of
              $\bv{\tilde{A}}$, \emph{i.e.}
              ${\widehat{\tau}_{i} \ge \bv{\tilde{a}}_i^{\top} (\bv{\tilde{A}}^{\top} \bv{\tilde{A}})^{\dagger}
              \bv{\tilde{a}}_i^{\top}},$
              such that $\sum_{i} \widehat{\tau}_{i} \leq 2n$.
                \item For every $i$, if $\beta \cdot \widehat{\tau}_{i} < 1,$ with
                  probability $\beta\cdot \widehat{\tau}_{i}$, divide
                  $\bv{\tilde{a}}_{i}$ by $\sqrt{\beta\cdot \widehat{\tau}_{i}}$; otherwise,
                  remove $\bv{\tilde{a}}_{i}$.
            \end{enumerate}
    \end{enumerate}
\item Return $\bv{\tilde{A}}$.
\end{enumerate}
\end{minipage}
}
\caption{The resparsifying streaming algorithm.}
\label{fig:streaming-sample}
\end{figure}

Consider a sparsification algorithm that reads edges of the graph one
by one and adds them to the sparsifier, and resparsifies when too many
edges have been accumulated.
Such an algorithm can be implemented in $\Oh(n \log n \epsilon^{-2})$ space and nearly linear time (see \Cref{fig:streaming-sample}).

\Cref{thm:main} gives that any such algorithm will end up
with a sparsifier of the original graph.
Some additional care is required to show that we can maintain a sparsifier of the current graph at all times.

\begin{lemma}
    \label{lem:stream-game}
    Let $\bv{\tilde{A}}$ be the matrix returned by $\streamSparsify(\bv{A}, \epsilon)$. Then, with high probability,
    \begin{align*}
        (1-\epsilon)\bv{A}^\top\bv{A} \preceq \bv{\tilde{A}}^\top\bv{\tilde{A}} \preceq (1+\epsilon)\bv{A}^\top\bv{A}.
    \end{align*}
\end{lemma}
\Proof To apply \Cref{thm:main}, we consider coupling the algorithm
with an adversary in a resparsification game. If the algorithm picks
an $i$ such that $\beta \cdot \widehat{\tau}_{i} < 1,$ and samples the row, the
adversary pick the same $i,$ and picks $p = \beta \widehat{\tau}_{i}.$ 
We argue by induction that the probability that the algorithm fails is
upper bounded by the probability that the adversary wins. 
At any step in the algorithm, assuming that the adversary has not won
so far, 
Since
$\widehat{\tau}_{i}$ is an upper bound on the leverage score, we have
\[
p 
\ge \beta \bv{\tilde{a}}_i^{\top} (\bv{\tilde{A}}^{\top}
\bv{\tilde{A}})^{\dagger} \bv{\tilde{a}}_i^{\top}
\ge {\beta} \bv{\tilde{a}}_i^{\top} \left(\sum_{i=1}^{j}
  \bv{\tilde{a}}_{i} \bv{\tilde{a}}_{i}^{\top} + \sum_{i=j+1}^{m}
  \bv{{a}}_{i} \bv{{a}}_{i}^{\top} \right)^{\dagger}
\bv{\tilde{a}}_i^{\top}
\ge \frac{\beta}{1+\eps}
\bv{\tilde{a}}_i^{\top} (\bv{{A}}^{\top} \bv{{A}})^{\dagger}
\bv{\tilde{a}}_i^{\top} \ge \alpha \bv{\tilde{a}}_i^{\top}
(\bv{{A}}^{\top} \bv{{A}})^{\dagger} \bv{\tilde{a}}_i^{\top},
\]
as required by the game. Hence, the adversary can make a legal move
identical to the algorithm.  Here, the third inequality uses the fact
that adversary has not won so far. If the adversary continues to play
the game till the end of the algorithm, the algorithm fails only if
the adversary wins before the last step. 

Thus, the algorithm succeeds with probability at least as much as the probability
that the adversary loses, which it does with high probability.
\QED

\Proofof{\Cref{thm:streaming}} For a graph $G(V,E)$ with edge weights
given by $w,$ we can write the Laplacian as
$\bv{L}_{G} = \sum_{e} (\sqrt{w_e} \bv{b}_{e}) (\sqrt{w_e}
\bv{b}_{e})^{\top},$
and hence apply~\cref{lem:stream-game}. It remains to show that step
3(b)i of \streamSparsify can be implemented in
$\Oh(n \beta \log^2 n)$ time.
This can be done by combining a fast solver for linear systems
in graph Laplacians~\cite{KoutisMP11,Peng13:thesis}, with
a Johnson-Lindenstrauss based estimation procedure,
as in~\cite{SpielmanS08, KoutisLP15,Peng13:thesis}.
\QED

%%% These are comments needed for working with emacs
%%% Local Variables: 
%%% mode: latex
%%% TeX-master: "resparsify"
%%% End: 

\section{Improved Combinatorial Algorithms
for (Parallel) Spectral Sparsification}
\label{sec:sparsify}

In~\cite{Koutis14}, Koutis gives a simple parallel
algorithm for graph sparsification.
Unlike other sparsification algorithms that rely on
effective resistances~\cite{SpielmanS08}, this algorithm
is combinatorial.
It finds, via a collection of spanners, upper bounds
of leverage scores that suffice for reducing edge counts.
The running time of this method, as well as a previous
spanner based sparsification algorithm by Kapralov
and Panigrahy~\cite{KapralovP12}  are several log
factors away from the numerical methods.
A summary of the best known bounds is presented in Figure~\ref{fig:previous}.

\begin{figure*}
\begin{center}
\begin{tabular}{c|c|c}
Algorithm / Reference & Sparsifier Size & Runtime (work)\\
\hline
Effective Resistances + Solvers ~\cite{SpielmanS08,KoutisMP11}
\footnotemark
& $\Oh(n\log{n} \epsilon^{-2})$
& $\Oh(m \log^2{n})$\\
\hline
Barrier functions~\cite{LeeS15} &
$\Oh(\theta n)$ & $\tilde{O}(m^{1 + 1/\theta})$\\
\hline
Graph partitioning~\cite{SpielmanT11,OrecchiaV11}
& $\Oh(n\log^6{n} \epsilon^{-2})$ & $\Oh(m\log^6{n})$\\
Random spanners~\cite{KapralovP12} & $\Oh(n\log^3{n}\epsilon^{-3})$ & 
$\Oh(m \log^{3}n \epsilon^{-3})$\\
Effective resistance sampling by spanners~\cite{Koutis14} &
$\Oh(n\log^3{n} \log^3(m/n) \epsilon^{-2}$
& $\Oh(m \log^2{n} \log^3(m / n) \epsilon^{-2})$\\
Resparsification (this paper) &
$\Oh(n\log^2{n} \log\log{n} \epsilon^{-2})$
& $\Oh(m \log^2{n} \epsilon^{-2})$\\
\end{tabular}

\caption{Runtime Bounds of Efficient Sparsification Algorithms}
\label{fig:previous}

\end{center}
\end{figure*}
\footnotetext{These bounds take into account the improvements
in~\cite{Peng13:thesis}.
The nearly $m\log^{1/2}$ time solvers in~\cite{CohenKMPPRX14}
are for vector-based guarantees.
Converting them into operator guarantees would incur
an additional factor of $\Oh(\log{n})$, leading to a higher
total than directly using the algorithm from~\cite{KoutisMP11}}

Compared to the algebraic sparsification algorithms,
all previous spanner based sparsification routines
have overheads of either $\log{n}$ or $\log\log^2{n}$
over the underlying spanner constructions.
This is due to the more gradual size reductions
offered by combinatorial routines for estimating
sampling probabilities.
Specifically, the spanner based routines from~\cite{Koutis14}
output a set of probabilities that upper
bound the true probabilities, but sum to
$\Oh(n h k) + m / k$
where $h$ is an overhead related to the combinatorial
algorithm and $k$ is a `reduction factor' that
can be picked.
As a result, such routines need to be invoked repeatedly
for up to $\Oh(\log{n})$ times to bring $m$ close to $n$.
This is handled either by doing some sampling ahead
of the estimation steps~\cite{KapralovP12},
or by having the errors compound during these steps~\cite{Koutis14}.
In each of these invocations, the $\Oh(n k h )$ term
is accumulated in the edge count, leading to an
overhead in edge count of at least $\log{n}$.
\footnote{It is possible to reduce this overhead
to $\poly{\log\log{n}}$ using steps similar to those
in~\cite{KoutisLP15}, but we're not aware of an explicit
statement of this.}

\begin{figure}[ht]
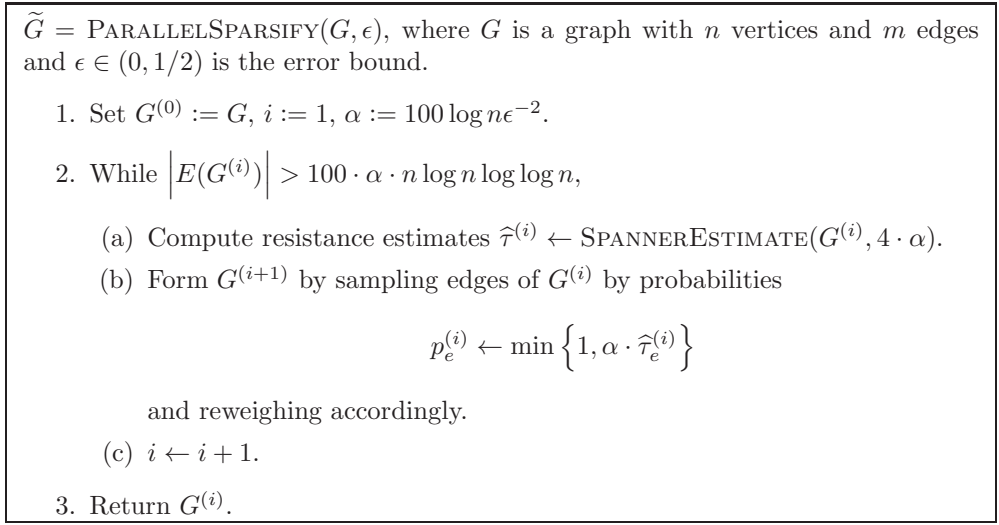

\noindent
\centering
\fbox{
\begin{minipage}{5in}
    \noindent $\widetilde{G} = \textsc{ParallelSparsify} (G, \epsilon)$,
    where $G$ is a graph with $n$ vertices and $m$ edges and
	$\epsilon \in (0, 1/2)$ is the error bound.
\begin{enumerate}
\item Set $G^{(0)} := G$, $i := 1$, $\alpha := 100 \log n \epsilon^{-2}$.
\item While $\abs{E(G^{(i)})} > 100 \cdot \alpha \cdot n\log n \log \log n$,
    \begin{enumerate}
        \item Compute resistance estimates
			$\widehat{\tau}^{(i)} \leftarrow \textsc{SpannerEstimate}(G^{(i)}, 
				4 \cdot \alpha)$.
		\item Form $G^{(i + 1)}$ by sampling edges of $G^{(i)}$ by probabilities
			\[
				p^{(i)}_e \leftarrow
					\min\left\{1, 
					\alpha \cdot \widehat{\tau}_{e}^{(i)}\right\}
			\]
            and reweighing accordingly.
			
		\item $i \leftarrow i + 1$.
    \end{enumerate}
\item Return $G^{(i)}$.
\end{enumerate}
\end{minipage}
}
\caption{Spectral Sparsification Algorithm}
\label{fig:sparsify}
\end{figure}

The guarantees we give for resparsification immediately imply
that these errors no longer accumulate.
Algorithmically it means it suffices to set the error
threshold in each of these steps to the desired final
accuracy, instead of one that's smaller by a factor of $\log{n}$.
This leads to the following result:
\begin{theorem}
\label{thm:sparsifyMain}
Given a graph $G$ with $n$ vertices, $m$ edges, and an
error parameter $\epsilon > 0$,
we can compute w.h.p. in $\Oh(m \log^2{n} \epsilon^{-2})$ work
and $\Oh(\epsilon^{-2} \log^4{n} \log^* n)$ depth a
$(1\pm\epsilon)$-spectral-sparsifier $H$ of $G$
with $\Oh(n\log^2{n} \log\log{n} \epsilon^{-2})$ edges.
\end{theorem}
Pseudocode of the algorithm is given in \Cref{fig:sparsify}.
It relies on a spanner based subroutine for
estimating total sampling probabilities.

In addition to this refined analysis of sparsification,
we also improve upon the leverage score estimation
from~\cite{Koutis14}.
Directly applying the parallel spanner algorithm
from~\cite{MillerPVX15} leads to a depth
dependent on $\log U$, where $U$ is the ratio
between the maximum and minimum edge weights in the graph.
In \Cref{subsec:spannerImproved}, we remove this
factor by relaxing the requirements needed for spanners.
This in fact further simplifies the algorithms.
In particular, we obtain an algorithm for estimating
leverage scores of the edges with the following guarantees:
\begin{lemma}
\label{lem:erBound}
  There exists a routine $\textsc{SpannerEstimate}$ that takes a
  graph $G$ with $n$ edges, $m$ vertices, and a parameter $\alpha \geq 1$ and computes in total work $\Oh(m \alpha \log^2 n \epsilon^{-2})$ and depth $\Oh(\log^3 n \epsilon^{-2} \log^* n)$
  estimates $\widehat{\tau}_e$ for all the edges such that
	with high probability:
	\begin{enumerate}
		\item For each edge $e$, we have
		$\tau_{e} \leq \widehat{\tau}_e \leq 1$ where
		$\tau_{e}$ is the true leverage score of $e$ in $G$.
        \item $ \sum_{e} \min\{1, \alpha\widehat{\tau}_e\}
			\leq \Oh(\alpha \cdot n \log n \log\log{n}) + m / 10$.
	\end{enumerate}
\end{lemma}

We use $\widehat{\tau}$ to denote the probability estimates
because $\tau$ is often used to denote exact statistical
leverage scores in the randomized numerical linear algebra
literature.
Such a routine then meets the properties of the
probability estimation algorithm
with $h = \Oh(\alpha \log{n} \log\log{n})$
and $k = 10$.
Applying the resparsification game as described
in Theorem~\ref{thm:main} with this routine
as the adversary then leads to the sparsification
algorithm shown in Figure~\ref{fig:sparsify}.

The correctness of this sparsification routine follows
from combining the estimation guarantees with the resparsification game.

\begin{proof}(of Theorem~\ref{thm:sparsifyMain})
It is easy to check that the resampling of edges in step 2b) of \textsc{ParallelSparsify} follows the rules of the game defined in \Cref{sec:game}, unless the game is already lost.
Thus, the fact that we obtain a sparsifier follows in a similar fashion to that in proof of \Cref{lem:stream-game}.

The edge count is given by the termination condition
given in Figure~\ref{fig:sparsify}, so it only remains
to bound running time.
By an ordinary Chernoff bound, we have that w.h.p. that as long as the edge count of $G^{(i)}$ is $\omega(n \log^2 n \log \log n \epsilon^{-2})$, it decreases by a constant factor in step 2b) with high probability.
As $m \leq n ^2$, this process terminates in $\Oh(\log{n})$ steps.

In each of these steps, the cost is dominated by the
computation of leverage score estimates
$\widehat{\tau}^{(i)}$ using \textsc{SpannerEstimate}
from Lemma~\ref{lem:erBound}.
As the edge counts are geometrically decreasing, this
total is within a constant factor of the first step,
giving a total cost of $\Oh(m \log^2{n} \epsilon^{-2})$.
\end{proof}

One further consequence of this leverage score oracle
is that the sparsifier is a union of forests.
This property is crucial to some uses of sparsifiers
in combinatorial algorithms, such as their recent
incorporation in data structures~\cite{AbrahamDKKP16:arxiv}.
By accounting for the structure of the output
of $\textsc{ProbabilisticSpanner}$ given in \Cref{subsec:spannerImproved}, we have a similar property.
\begin{corollary}
The output of Theorem~\ref{thm:sparsifyMain} can be
written as a sum of $\Oh(\log^{2}n \log\log{n} \epsilon^{-2})$ forests.
\end{corollary}
This result is similar to the ones from~\cite{ChuCPP14},
and improves the dynamic sparsification algorithms
in~\cite{AbrahamDKKP16:arxiv} by a factor of $\log^2{n}$.
Furthermore, its running time is within $\Oh(\log\log{n})$ factor
of numerically oriented routines based on random projections
and solving linear systems~\cite{SpielmanS08}.
As a result, we're optimistic about the
practical potential of this sparsification approach.

\Authornote{Richard}{I tried to rewrite this a bit,
it's even vaguer now, but the main point that I tried
to get across is $\Oh(m + n\log^2{n})$ is the `right'
bound to aim for.}
On the other hand, this improvement over~\cite{Koutis14}
translates to a fairly minor improvement for constructing
sparsifier chains~\cite{Koutis14,ChengCLPT15,KyngLPSS16}.
This is because the running time of those routines directly
depend on running the sparsification algorithm on its own output.
In our case, as the runtime depends on $m$,
the two factors of $\log^2{n}$ accumulate to $\log^4{n}$,
a rather large overhead.
Further improvements in this direction would require lowering
the dependencies in $m$, in ways similar to the 
$\Oh(m + n\log^c n)$ time sparsification algorithms in 
in~\cite{KoutisLP15,FungHHP11,HariharanP10:arxiv,JindalK15:arxiv}.

\subsection{Faster Probabilistic Spanners}
\label{subsec:spannerImproved}
Koutis~\cite{Koutis14} obtained estimates
of effective resistance estimates as needed in
Lemma~\ref{lem:erBound} through combinatorial
combinations of resistors in series and parallel.
Here the resistance of a single resistor
can be viewed as the inverse of its weight.
Low resistance estimates can be obtained via
the following two facts:
\begin{fact}
\label{fact:rSP}
\leavevmode
\begin{enumerate}
\item \label{part:rSeries}
A sequence of resistors in series with resistances
$r_1, r_2, \ldots, r_k$ has effective resistance
$r_1 + r_2 + \ldots + r_k$.
\item \label{part:rParallel}
A set of $k$ parallel, non-overlapping paths,
each with effective resistance at most $r$, has resistance at most $r / k$.
\end{enumerate}
\end{fact}
Part~\ref{part:rSeries} suggests that short paths
can give reasonable bounds, while Part~\ref{part:rParallel}
suggests that these bounds can be further improved
by having multiple disjoint paths.
Koutis~\cite{Koutis14} combines these by repeatedly removing
spanners from a graph.
Each spanner is a sparse graph that guarantees that all
remaining edges have `shortcuts' that are longer by
a factor of $\Oh(\log{n})$.
Specifically, it guarantees that the stretch of an edge $e = uv$,
\[
str_H\left(e = uv\right) \defeq \frac{dist_H\left(u, v\right)}{l_{e}},
\]
where $dist_H\left(u, v\right)$ is the shortest path
distance in $l$ in $H$ between $u$ and $v$,
is bounded by $\Oh(\log{n})$.

If the lengths are set to resistances, specifically
$l_e = \frac{1}{w_e}$, each spanner provides a
path of resistance $\frac{\Oh(\log{n})}{w_e}$ meeting
condition Part~\ref{part:rSeries} for the edge $e$.
Removing this spanner and building another one then
gives another such path, and repeating then gives
the $k$ disjoint paths need in Part~\ref{part:rParallel}.
Our primary improvement is demonstrating that
\emph{probabilistic} spanners also suffice for these purposes.

\begin{Definition}
    The random graph $H$ is a \emph{probabilistic $k$-spanner} for $G$ if $H \subseteq G$ and every edge in $G$
	has probability at least $1/2$ to have stretch at
	most $k$ in $H$.
\end{Definition}

In the rest of this section, we show that probabilistic
$\Oh(\log n)$-spanners of weighted graphs of size
$\Oh(n \log \log n)$ can be constructed in $\Oh(m)$
work and $\Oh(\log n \log^* n)$ parallel depth.
Compared to the parallel spanner construction
from~\cite{MillerPVX15}, our routine has depth
that's lower by a factor of $\log{U}$, where
$U$ is the ratio of the maximum to minimum edge weights.
Our algorithm is based on directed invocations of an
exponential start time clustering routine, whose
guarantees as given
in~\cite{MillerPX13} and~\cite{MillerPVX15}
\footnote{We omit a precise pointer as these manuscripts
are still undergoing edits.} are:
\begin{lemma}
\label{lem:estCluster}
There is a routine $\textsc{ESTCluster}$ that
given a graph $G = (V, E)$ with $n$
vertices, $m$ edges, and all weights $w_e \geq 1$, along
with a parameter $\beta > 0$,
$\textsc{ESTCluster}(G, \beta)$ generates
in $\Oh(\beta^{-1} \log n \log^* n)$ depth and $\Oh(m)$ work
a partition of the vertices into clusters
$\mathcal{X} = (X_1 \cup X_2 \cup \ldots \cup X_k)$ such that:
\begin{enumerate}
\item $X_1 \ldots X_k$ form a disjoint partition of $V$.
\label{part:partition}
\item With high probability, the combinatorial (unweighted)
diameter of each $X_i$ is certified by a spanning
tree on $X_i$ with diameter $\Oh(\beta^{-1} \log{n})$.
\label{part:diam}
\item For any neighboring vertices $u$ and $v$,
the probability that $u$ and $v$ are in
different clusters is at most $\beta$.
\label{part:cutProb}
\end{enumerate}
\end{lemma}

Although this routine can be extended to weighted
graphs, spanner constructions using it also
work with the unweighted variant~\cite{MillerPVX15}.
We will invoke it by first partitioning up the graphs by
weight scales (which we denote with superscripts $^{(i)}$),
applying this clustering routine to each scale separately,
and then combining the results using minimum spanning trees.
Pseudocode of this algorithm is given in Figure~\ref{fig:prob-spanner}.

\begin{figure}[ht]
\noindent
\centering
\fbox{
\begin{minipage}{5in}
    \noindent $S = \textsc{ProbSpanner} (G, l)$,
    where $G$ is a graph with $n$ vertices and $m$ edges,
	and $l$ are lengths on the edges.
\begin{enumerate}
\item For integers $i$ in parallel:
\begin{enumerate}
\item Let $E^{(i)}$ be the set of edges in $G$ of length
in $[2^i, 2^{i+1})$,
and $G^{(i)}$ be the unweighted graphs formed from $E^{(i)}$.
\item Let $\mathcal{X}^{(i)}
\leftarrow \textsc{ExpCluster}(G^{(i)}, 1/3)$,
and $C^{(i)}$ be the
union of the BFS trees certifying the diameters
of the clusters in $\mathcal{X}^{(i)}$, but with
the original edge weights from $G$.
\end{enumerate}
\item Let $t := \Oh(\log \log n)$.
\item  \label{step:MST} For each $0 \leq j < t$,
let $F^{(j)}$ be a minimum spanning forest of $\bigcup_{i \equiv j \pmod{t}} C^{(i)}$.
\item Return the union of $F^{(j)}$.
\end{enumerate}
\end{minipage}
}
\caption{Probabilistic Spanner Algorithm}
\label{fig:prob-spanner}
\end{figure}

\begin{theorem}
    The random graph
	$S := \textsc{ProbabilisticSpanner}(G)$
	is a probabilistic $\Oh(\log n)$-spanner of $G$.
    Moreover, it is the sum of $\Oh(\log \log n)$ trees,
	and can be computed in $\Oh(m)$ total work and
	$\Oh(\log^2 n \log^* n)$ depth.
\end{theorem}

\Proof
The work and depth bounds follow from Lemma~\ref{lem:estCluster}
and the cost of finding minimum spanning trees in
parallel~\cite{Jaja92:book,Leighton92:book}.
So it suffices to show that each edge has stretch
$\Oh(\log{n})$ with probability at least $1/2$.

Consider an edge $e = uv \in E^{(i)}$.
Part~\ref{part:cutProb} of Lemma~\ref{lem:estCluster}
gives that the probability that $u$ and $v$
belong to the same cluster of $\mathcal{X}^{(i)}$
with probability at least $1 - \beta = \frac{2}{3}$.
Combining this with Part~\ref{part:diam} of
Lemma~\ref{lem:estCluster} gives that with probability
at least $1/2$ we have that $e$ is contained in some
cluster $X^{(i)}_{k}$, and that the combinatorial
diameter of all clusters is $c_0 \cdot \log{n}$
for some absolute constant $c_0$.
We will now use these two conditions to bound the
stretch of $e$ in $F^{(j)}$, where $0 \leq j < t$
and $j \equiv i \pmod{t}$.

Let $i_1 < i_2 < \ldots < i_k$ be the indices
$\equiv i \pmod{t}$ for which $E^{(i)}$ is nonempty.
We will show by induction that if
$2^{t} \geq 8 c_0 \cdot \log{n}$,
then the diameter of any connected component
of any minimum spanning forest $T^{(q)}$ of
$C^{(i_1)} \cup \ldots \cup C^{(i_q)}$ is bounded by
\[
	4 c_0 \cdot 2^{i_q} \log{n}.
\]

The base case of $q = k$ follows directly
from the diameter bound.

For the inductive case, the inductive hypothesis
gives that all connected components of $T^{(q - 1)}$
have diameter at most $4 c_0 \cdot 2^{i_q} \log{n}$,
while we also have that the length of each
edge in $E^{(i_q)}$ is at most $2^{i_q + 1}$,
Since the combinatorial diameter of each connected
component of $C^{(i_q)}$ is $c_0 \log{n}$, the
diameter of $T^{(q)}$ is bounded by the lengths
of these edges / components alternating, giving:
\[
   c_0 \log{n} \cdot 2^{i_q + 1}
	+ \left(c_0 \log{n} + 1\right)
	\left( 4 c_0 \cdot 2^{i_{q - 1}} \log{n} \right).
\]
The condition of $2^{t} \geq 8 c_0 \log{n}$
and $i_{q} \geq i_{q - 1} + t$ then imply
$2^{i_{q}} \geq 2^{i_{q - 1}}$, and the
total is bounded by $4 c_0 2^{i_q} \log{n}$,
so the inductive hypothesis holds for $q$ as well.

The bound on stretch then follows from
$e$ having length at least $2^{i}$.
\QED

\begin{figure}[ht]
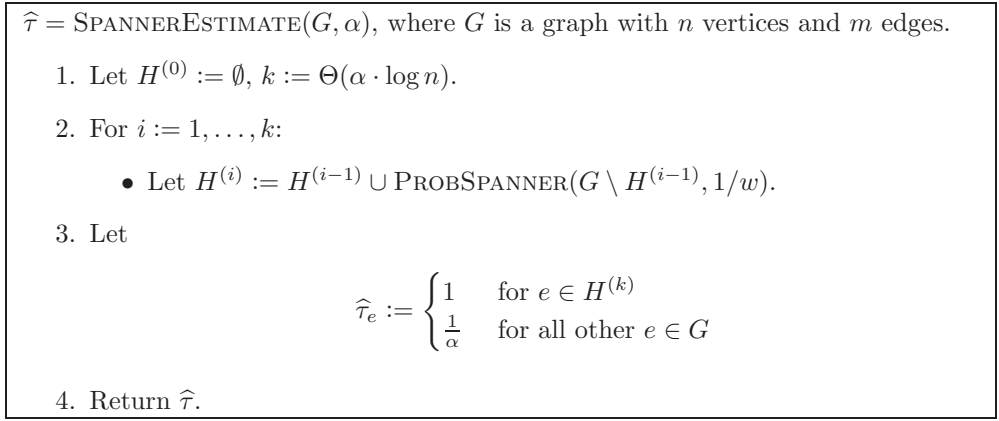

\noindent
\centering
\fbox{
\begin{minipage}{5in}
    \noindent $\widehat{\tau} = \textsc{SpannerEstimate} (G, \alpha)$,
    where $G$ is a graph with $n$ vertices and $m$ edges.
\begin{enumerate}
\item Let $H^{(0)} := \emptyset$, $k :=  \Theta( \alpha \cdot \log n )$.
\item For $i := 1, \ldots, k$:
    \begin{itemize}
        \item Let $H^{(i)} := H^{(i-1)} \cup \textsc{ProbSpanner}(G \setminus H^{(i-1)}, 1/w)$.
    \end{itemize}
\item Let
    \begin{align*}
        \widehat{\tau}_e := \begin{cases}
            1 & \mbox{ for $e \in H^{(k)}$}\\
            \frac{1}{\alpha} & \mbox{ for all other $e \in G$}\\
        \end{cases}
    \end{align*}
\item Return $\widehat{\tau}$.
\end{enumerate}
\end{minipage}
}

\caption{Probabilistic Spanner Based Leverage Score Estimation}
\label{fig:spanner-estimate}
\end{figure}

Repeatedly invoking this probabilistic spanner
routine then leads to the leverage score
estimation algorithm.
Its pseudocode is given in Figure~\ref{fig:spanner-estimate}.

\Proofof{\Cref{lem:erBound}}
The work and depth guarantees follow from the cost of invoking \textsc{ProbabilisticSpanner} $\Oh(\alpha \cdot \log n)$ times.

For the stretch bounds, by Chernoff bounds for scalars,
since $k \geq \Omega(\log{n})$, we have that with high
probability every remaining edge will have stretch at
most $\Oh(\log{n})$ w.r.t. at least $k / 3$ of the
removed spanners.
This means that for every remaining edge $e$
there are $\Omega(\alpha \cdot \log{n})$
edge disjoint paths each of which has length at most $\Oh(\log{n})$
times the length of $e$.
Since the length of each edge was set to its resistance,
 Fact~\ref{fact:rSP} then gives that these paths upper
bound the statistical leverage score of the edge by
$1/(10 \alpha)$.
As the total number of edges in each spanner is $\Oh(n\log\log{n})$,
we get $ \sum_{e} \min\{1, \alpha\widehat{\tau}_e\}
			\leq \Oh(\alpha \cdot n \log n \log\log{n}) + m / 10$.
\QED

%%% These are comments needed for working with emacs
%%% Local Variables: 
%%% mode: latex
%%% TeX-master: "resparsify"
%%% End: 

\FloatBarrier
\section*{Acknowledgments}
The authors would like to thank Michael Cohen, Yiannis Koutis, and
Shen Chen Xu for helpful comments and discussions.
Rasmus Kyng is partially supported by ONR Award N00014-16-1-2374 and 
NSF Grant CCF-1111257.
Jakub Pachocki did part of this work while a student at Carnegie
Mellon University is partially supported by the ONR under
Grant No. N00014-15-1-238 and the NSF under Grant No. 1065106.
Richard Peng is partially supported by the NSF under Grant No. 1637566.
Sushant Sachdeva was a post-doc at the Department of Computer Science,
Yale University when this work was done. He was
supported by a Simons Investigator Award to Daniel A. Spielman.

\bibliographystyle{alpha}
\bibliography{papers}

\end{document}